\documentclass[prl,showpacs,superscriptaddress,twocolumn]{revtex4}

\usepackage{amsmath}
\usepackage{graphicx}

\newcommand{\cordoba}{Facultad de Matem\'{a}tica, Astronom\'{\i}a y
F\'{\i}sica, Universidad Nacional de C\'{o}rdoba, 5000 C\'{o}rdoba, Argentina}
\newcommand{\strasbourg}{Institut de Physique et Chimie des
Mat\'{e}riaux de Strasbourg, UMR 7504, CNRS-ULP,\\
23 rue du Loess, BP 43, 67034 Strasbourg Cedex 2, France}

\begin{document}

\title{Semiclassical Theory of Time-Reversal Focusing}

\author{Hern\'an~L. Calvo}
\affiliation{\cordoba}
\affiliation{\strasbourg}
\author{Rodolfo~A.~Jalabert}
\affiliation{\strasbourg}
\author{Horacio~M.~Pastawski}
\affiliation{\cordoba}

\begin{abstract}
Time reversal mirrors have been successfully implemented for various kinds of
waves propagating in complex media. In particular, acoustic waves in chaotic
cavities exhibit a refocalization that is extremely robust against external
perturbations or the partial use of the available information. We develop a
semiclassical approach in order to quantitatively describe the refocusing
signal resulting from an initially localized wave-packet. The time-dependent
reconstructed signal grows linearly with the temporal window of injection, in
agreement with the acoustic experiments, and reaches the same spatial
extension of the original wave-packet. We explain the crucial role played by
the chaotic dynamics for the reconstruction of the signal and its stability
against external perturbations.

\end{abstract}

\maketitle

The concept of time reversal has captured the imagination of physicists for
more than a century, leading to a vast theoretical oeuvre, sempiternal
discussions, and a few concrete experimental realizations. Among them, the
works on spin echoes have been of paramount importance concerning the limits
in the reconstruction of an initially prepared quantum state
\cite{cit-Hahn,cit-Zhang}. The time reversal of acoustic waves in a
non-homogeneous medium was another experimental deed showing that an initially
localized pulse can be accurately reconstructed by an array of
receiver-emitter transducers that re-inject the recorded signal
\cite{cit-Derode}. Re-focusing of elastic, as well as electromagnetic, waves
has been later achieved \cite{cit-Draeger,cit-Catheline,cit-Lerosey}. These
experiments provoke a natural surprise while yielding reconstructions, that
albeit not perfect, are highly faithful. The relevant questions that arise
when trying to understand these physical realizations of time reversal are
related with how good a reconstruction can be achieved and which are the
limits set by interactions with the environment and unavoidable errors in the
reversal protocol.

In the spin echo experiments the complexity of the physical system has emerged
as a critical component, and the term of Loschmidt echo (LE) has been coined
to describe setups where many-body physics or chaotic dynamics are relevant
\cite{cit-Levstein}. In particular, the decay of the LE with the reversal time
has been shown to depend on the underlying classical dynamics
\cite{cit-Peres,cit-Jalabert,cit-Gorin}: classically chaotic systems exhibit a
decay rate which, for large perturbations, is bounded by the Lyapunov exponent
characterizing the dynamics.

In the time-reversal mirror (TRM) procedure the play back signal builds up in
the region of the original excitation, in the form of a reversed wave
amplitude \cite{cit-Derode}. Thus, the TRM can be viewed as the wave version
of the LE. A salient feature of the TRM experiments is that, even though
reversal is not perfect \cite{cit-Pastawski}, the refocusing improves when the
wave-propagation occurs in a disordered medium or in a chaotic cavity, as
compared with the homogeneous or integrable case. Remarkably, a single
transducer is enough in the case of a chaotic cavity \cite{cit-Draeger}. The
asymptotic analysis of the Wigner transform of wave fields in the
high-frequency limit has been used to understand how multiple scattering
enhances the spatial resolution of the refocused signal \cite{cit-Bal}.
Diagrammatic perturbation theory has been able to account for the
symmetry-induced interference enhancements in the refocalization observed in
disordered media \cite{cit-deRosny}. The refocusing experiments in chaotic
cavities have been confronted with numerical simulations \cite{cit-Draeger},
as well as ergodicity and control theory \cite{cit-Bardos}. The contrasting
stability properties of TRM with wave and particle propagation through a
multiple scattering medium has been discussed in Ref. \cite{cit-snie}.

In this work we develop a semiclassical approach for TRM in chaotic cavities
and quantify the quality of the reconstructed signal in terms of temporal and
spatial dispersions, as well as possible environmental influences. We
demonstrate the crucial role played by the underlying classical dynamics and
validate our analytical results by confronting them to numerical simulations.

\begin{figure}[ptb]
\includegraphics[width=8.6cm]{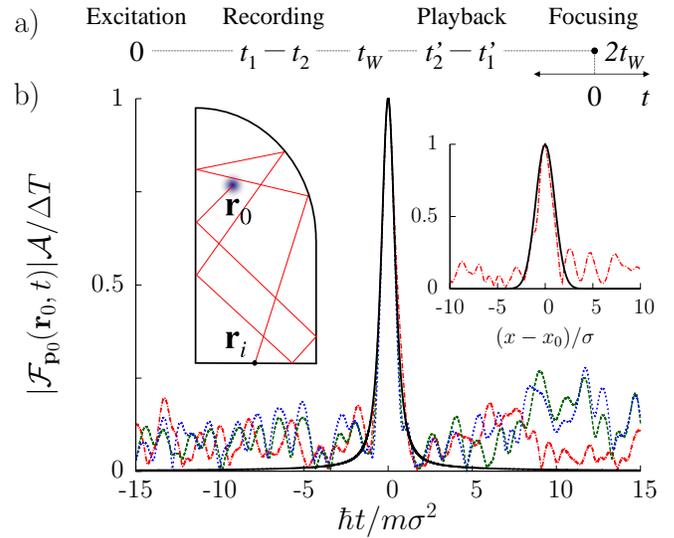}\caption{(color online). (a) TRM
sequence. (b) Reconstructed signal scaled with $\mathcal{A}/\Delta T$ at the
emission point $\mathbf{r}_{0}$, close to the refocusing time $t=0$, for the
shown billiard. The thick solid is the semiclassical prediction
(Eq.~(\ref{eq:F_p0})). Numerical simulations for various $\Delta T$ and $\mathcal{A}$:
$5000$ and $150\times300$ (blue dotted), $10000$ and $150\times300$ (green
dashed), and $5000$ and $300\times600$ (red dash-dotted). Right inset:
reconstructed signal at $t=0$ close to $\mathbf{r}_{0}$ from the semiclassical
prediction (solid) and simulation (red dash-dotted) with $\Delta T=5000$ and
$\mathcal{A}=300\times600$.}%
\label{fig-1}%
\end{figure}

A high-frequency signal emitted at $t=0$ at a position $\mathbf{r}_{0}$ inside
the cavity can be interpreted within the ray picture as an initial wave-packet
centered at $\mathbf{r}_{0}$ that evolves and is recorded by a receiver (or an
array of receivers) at position(s) $\mathbf{r}_{i}$ ($i=1,2,\ldots,N$) for
times in the interval $(t_{1},t_{2})$. After a waiting time $t_{W}>t_{2}$ the
re-emission of the time-reversed signal is performed in the interval
$(t_{2}^{\prime}=2t_{W}-t_{2},t_{1}^{\prime}=2t_{W}-t_{1})$. The refocusing is
expected at $2t_{W}$, that is redefined as the time origin for refocusing
\cite{cit-Draeger} (see Fig.~\ref{fig-1}(a)). The signal that can be detected
in a point $\mathbf{r}$ at a time $t$ is \cite{cit-Tanner}%
\begin{align}
\mathcal{F}_{\mathbf{p}_{0}}(\mathbf{r},t)=&\sum_{i}\int_{t_{1}}^{t_{2}%
}\mathrm{d}\tau\ G(\mathbf{r,r}_{i},t+\tau)\nonumber\\
& \times\int\mathrm{d}\mathbf{r}^{\prime}\ G^{\ast}(\mathbf{r}_{i}%
,\mathbf{r}^{\prime},\tau)\psi_{\mathbf{p}_{0}}^{\ast}(\mathbf{r}^{\prime
})\ .\label{eq_twoGF}%
\end{align}
The propagator $G(\mathbf{r},\mathbf{r}_{i},t+\tau)$ corresponds to the
re-emitted signal, which is obtained by time-reversing the evolution of the
initial state with the propagator $G(\mathbf{r}_{i},\mathbf{r}^{\prime},\tau
)$. We do not write the initial temporal arguments of the propagators, as they
are taken to be $0$. We work in two dimensions and choose as an initial state
a Gaussian wave-packet%
\begin{equation}
\psi_{\mathbf{p}_{0}}(\mathbf{r}^{\prime})=\frac{1}{\sqrt{\pi}\sigma}%
\exp\left[  \tfrac{\mathrm{i}}{\hbar}\mathbf{p}_{0}\cdot({\mathbf{r}}^{\prime
}-\mathbf{r}_{0})-\tfrac{(\mathbf{r}^{\prime}-\mathbf{r}_{0})^{2}}{2\sigma
^{2}}\right]  \ ,
\end{equation}
centered around $\mathbf{r}_{0}$ and with dispersion $\sigma$. The initial
momentum $\mathbf{p}_{0}$ sets the energy and direction of the original
signal. The choice of a quantum formalism to represent the ray picture is
motivated by convenience, as we are leaving aside the delicate issue
concerning a quantal recording-emission process. The quantum formalism is
suited to work with the semiclassical propagator \cite{cit-Brack}%
\begin{equation}
G(\mathbf{r}^{\prime\prime},\mathbf{r}^{\prime},\tau)=\frac{1}{2\pi
\mathrm{i}\hbar}\sum_{s(\mathbf{r}^{\prime},\mathbf{r}^{\prime\prime},\tau
)}\sqrt{C_{s}}\ e^{\mathrm{i}S_{s}/\hbar-\mathrm{i}\frac{\pi}{2}\mu_{s}%
},\label{eq_semiclassicalGF}%
\end{equation}
given as a sum over classical trajectories $s$ traveling in a time $\tau$
between the two extreme points. We note $S_{s}=S_{s}(\mathbf{r}^{\prime\prime
},\mathbf{r}^{\prime},\tau)$ the action integral along the path, $\mu_{s}$ the
Maslov index, and $C_{s}=|\mathrm{det}(-\partial S_{s}/\partial\mathbf{r}%
^{\prime}\partial\mathbf{r}^{\prime\prime})|$.

Performing the $\mathbf{r}^{\prime}$-integral of Eq.~(\ref{eq_twoGF}) by
stationary-phase (see Ref.~\cite{cit-Goussev} for the precise conditions of
validity of such an approximation in a chaotic cavity) we can write, for a
single transducer,%
\begin{align}
\mathcal{F}_{\mathbf{p}_{0}}(\mathbf{r},t)=&\frac{\sigma}{2\pi^{3/2}%
\hbar^{2}}\int_{t_{1}}^{t_{2}}\mathrm{d}\tau\sum_{s^{\prime}(\mathbf{r}%
_{i},\mathbf{r,}\tau+t)}\sum_{s(\mathbf{r}_{0},\mathbf{r}_{i},\tau)}%
\sqrt{C_{s^{\prime}}C_{s}}\nonumber\\
& \times\exp\left[  \tfrac{\mathrm{i}}{\hbar}(S_{s^{\prime}}-S_{s}%
)-\mathrm{i}\tfrac{\pi}{2}(\mu_{s^{\prime}}-\mu_{s})\right. \nonumber\\
& \left.  -\tfrac{\sigma^{2}}{2\hbar^{2}}(\mathbf{p}_{s}-\mathbf{p}_{0}%
)^{2}\right]  \ ,\label{eq_twoGF_semi}%
\end{align}
where $\mathbf{p}_{s}$ is the initial momentum of the trajectory $s$.

We are interested in times $t$ close to the refocusing one, and positions
$\mathbf{r}$ near $\mathbf{r}_{0}$, therefore the dominating contribution
comes from the diagonal approximation $s^{\prime}\simeq s$ leading to a signal
given by
\begin{subequations}
\label{eq:allTGF}%
\begin{align}
\mathcal{F}_{\mathbf{p}_{0}}(\mathbf{r},t)=&\frac{\sigma}{2\pi^{3/2}%
\hbar^{2}}\int_{t_{1}}^{t_{2}}\mathrm{d}\tau\ F_{\mathbf{p}_{0}}%
(\mathbf{r}_{0},\mathbf{r}_{i},\mathbf{r},\tau),\label{eq:TGF0}\\
\displaystyle F_{\mathbf{p}_{0}}(\mathbf{r}^{\prime},\mathbf{r}^{\prime\prime
},\mathbf{r},\tau)=&\sum_{s(\mathbf{r}^{\prime},\mathbf{r}^{\prime\prime
},\tau)}C_{s}\ f_{\mathbf{p}_{0}}(\mathbf{r}^{\prime},\mathbf{r}^{\prime
\prime},\mathbf{r},\mathbf{p}^{\prime}),\label{eq:TGF1}\\
\displaystyle f_{\mathbf{p}_{0}}(\mathbf{r}^{\prime},\mathbf{r}^{\prime\prime
},\mathbf{r},\mathbf{p}^{\prime})=&\exp\left[  \tfrac{\mathrm{i}}{\hbar
}\mathbf{p}^{\prime}\cdot(\mathbf{r}-\mathbf{r}^{\prime})\right. \nonumber\\
&  \left.  -\tfrac{\mathrm{i}}{\hbar}E_{s}t-\tfrac{\sigma^{2}}{2\hbar^{2}%
}(\mathbf{p}^{\prime}-\mathbf{p}_{0})^{2}\right]  .\label{eq:TGF2}%
\end{align}
$E_{s}$ is the energy at which the trajectory $s$ is traveled. In billiards
the magnitude of the momentum modifies the traveling time but does not affect
the path. Noting $\hat{s}(\mathbf{r}_{0},\mathbf{r}_{i})$ the geometrical
support of the trajectory $s(\mathbf{r}_{0},\mathbf{r}_{i},\tau) $ with length
$L_{\hat{s}}$, we have $E_{s}=\mathrm{p}_{s}^{2}/2m$ and $\mathrm{p}%
_{s}=(mL_{\hat{s}}/\tau)$, where $m$ is the mass of the particle.

In order to present the calculation in its simplest form we start by setting
$\mathbf{p}_{0}=0$ and the optimal conditions $t=0$ and $\mathbf{r}%
=\mathbf{r}_{0}$, which yield (from the last term in the exponent of
(\ref{eq:TGF2})) the maximum refocusing $\mathcal{F}_{\mathrm{max}%
}=\mathcal{F}_{0}(\mathbf{r}_{0},0)$. In a fully chaotic system, $C_{s}$
scales as $\exp[-\lambda_{s}\tau]$, where $\lambda_{s}$ is the largest
Lyapunov exponent. Assuming further a uniformly hyperbolic dynamics
\cite{cit-Goussev} and using that in a billiard $\lambda_{s}\tau={\hat
{\lambda}}L_{\hat{s}}$ (with ${\hat{\lambda}}$ as an inverse length), we write
$C_{s}=2m^{2}{\hat{\lambda}}L_{\hat{s}}/\tau^{2}\exp[-{\hat{\lambda}}%
L_{\hat{s}}]$. Noting $\mathcal{V}^{2}=2\hbar^{2}/m^{2}\sigma^{2}$ we have
\end{subequations}
\begin{equation}
\mathcal{F}_{\mathrm{max}}=\frac{\sigma m^{2}{\hat{\lambda}}}{\pi^{3/2}%
\hbar^{2}}\sum_{\hat{s}(\mathbf{r}_{0},\mathbf{r}_{i})}L_{\hat{s}}%
e^{-\hat{\lambda}L_{\hat{s}}}\int_{t_{1}}^{t_{2}}\frac{\mathrm{d}\tau}%
{\tau^{2}}\exp\left[  -\left(  \frac{L_{\hat{s}}}{\mathcal{V}\tau}\right)
^{2}\right]  .\label{eq_chi0}%
\end{equation}

The sum over the transient orbits $\hat{s}$ can be converted into an integral
over trajectory lengths by introducing the density $\mathrm{d}N(L)/\mathrm{d}%
L=\pi/(\hat{\lambda}\mathcal{A})\exp{(\hat{\lambda}L)}$ ($\mathcal{A}$ stands
for the area of the chaotic cavity) \cite{cit-Sieber99,cit-Goussev-priv}.
Denoting $L_{j}=\mathcal{V}t_{j}$ ($j=1,2$) and $L_{\mathrm{d}}$ the length of
the shortest trajectory linking $\mathbf{r}_{0}$ and $\mathbf{r}_{i}$ we have%
\begin{align}
\mathcal{F}_{\mathrm{max}}=&\frac{1}{\sigma\mathcal{A}}\left\{
(t_{2}-t_{1})\int_{L_{\mathrm{d}}/L_{2}}^{\infty}\mathrm{d}l\left[
1-\mathrm{erf}\left(  l\right)  \right]  \right. \nonumber\\
& \left.  +t_{1}\int_{L_{\mathrm{d}}/L_{2}}^{L_{\mathrm{d}}/L_{1}}%
\mathrm{d}l\left[  1-\mathrm{erf}\left(  l\right)  \right]  \right\}
\ ,\label{eq:F_m-inter}%
\end{align}
where $\mathrm{erf}(x)$ stands for the error function. The assumptions made on
$C_{s}$ and $\mathrm{d}N(L)/\mathrm{d}L$ are valid for lengths a few times
larger than $L_{\mathrm{d}}$. However, our approximation is appropriate since
we assume that we start recording at times $t_{1}$ large enough for the
typical contributing trajectories to feel the chaotic nature of the dynamics.
That is, we work under the hypothesis $L_{\mathrm{d}}\ll L_{1}<L_{2}$, that
also allows to neglect the last integral, leading to%
\begin{equation}
\mathcal{F}_{\mathrm{max}}=\frac{1}{\sqrt{\pi}\sigma\mathcal{A}}\ \Delta
T\ .\label{eq:F_m}%
\end{equation}
The scaling of the refocused signal with the injection interval $\Delta
T=t_{2}-t_{1}$ is a quite natural result, experimentally observed in
Ref.~\cite{cit-Draeger}, while the scaling with $\mathcal{A}$ has not been
systematically tested so far. In the case where there is an array with $N$
transducers, we simply have to multiply the above result by $N$, but the
surprising fact that just one detector is enough stems from Eq.~(\ref{eq:F_m}%
). In order to further quantify faithfulness of the time-reversal process we
need to evaluate the temporal and spatial extents of the reconstructed signal.

For times $t$ close to the refocusing one we have to consider the phase
$mL_{\hat{s}}^{2}t/(2\tau^{2}\hbar)$ in Eq.~(\ref{eq:TGF2}). Therefore,
$\mathcal{F}_{0}(\mathbf{r}_{0},t)$ follows the same expression than
$\mathcal{F}_{\mathrm{max}}$ in Eq.~(\ref{eq_chi0}) by the change of
$1/\mathcal{V}^{2}$ into $1/\mathcal{V}^{2}+\mathrm{i}mt/2\hbar$. The error
functions resulting from the $\tau$-integral have now to be extended into the
complex plane, and after a straightforward algebra we find
\begin{equation}
\mathcal{F}_{0}(\mathbf{r}_{0},t)=\frac{\mathcal{F}_{\mathrm{max}}%
}{1+\mathrm{i}\mathcal{V}t/(\sqrt{2}\sigma)}\ .\label{eq:F_time}%
\end{equation}

The reader can imagine how the general calculation goes when we treat
simultaneously $t\neq0$, $\mathbf{r}\neq\mathbf{r}_{0}$ and $\mathbf{p}%
_{0}\neq0$. Instead of presenting such calculation, which results in the
faithful reconstruction of the initial wave-packet, we look at the problem
from a different perspective and introduce the ergodicity hypothesis in order
to treat the general case. The ergodic approach not only provides a second,
and more economical, way of obtaining the general result without using
detailed knowledge of the dynamics, but also sheds some light into the
necessary conditions for achieving the refocalization condition. The basics of
the ergodic approach is to calculate quantities like $F_{\mathbf{p}_{0}%
}(\mathbf{r}^{\prime},\mathbf{r}^{\prime\prime},\mathbf{r},\tau)$ of
Eq.~(\ref{eq:TGF1}) by averages over phase-space \cite{cit-Argaman}. Calling
$\mathbf{r}_{\tau}=\mathbf{r}_{\tau}(\mathbf{r}^{\prime},\mathbf{p}^{\prime})$
and $\mathbf{p}_{\tau}=\mathbf{p}_{\tau}(\mathbf{r}^{\prime},\mathbf{p}%
^{\prime})$ the position and momentum, respectively, at time $\tau$ of a
particle starting at $(\mathbf{r}^{\prime},\mathbf{p}^{\prime}) $ at time $0$,
we have%
\begin{align}
F_{\mathbf{p}_{0}}(\mathbf{r}^{\prime},\mathbf{r}^{\prime\prime}%
,\mathbf{r},\tau)=&\int\mathrm{d}\mathbf{p}^{\prime} \mathrm{d}%
\mathbf{p}^{\prime\prime}\delta\left(  \mathbf{r}_{\tau}-\mathbf{r}%
^{\prime\prime}\right)  \delta\left(  \mathbf{p}_{\tau}-\mathbf{p}%
^{\prime\prime}\right) \nonumber\\
& \times f_{\mathbf{p}_{0}}(\mathbf{r}^{\prime},\mathbf{r}^{\prime\prime
},\mathbf{r},\mathbf{p}^{\prime}).
\end{align}

The double delta-function represents the distribution of classical
trajectories. An average over small ranges of initial and final conditions
gives a smooth distribution which describes the evolution in a statistical
sense. For sufficiently long times such a distribution is $\tau$ independent,
and uniformly distributed on the hyper-surface of constant energy (which for
two dimensional billiards has a volume $\Omega=2\pi m\mathcal{A}$ in
phase-space). We therefore have%
\begin{align}
F_{\mathbf{p}_{0}}(\mathbf{r}^{\prime},\mathbf{r}^{\prime\prime}%
,\mathbf{r},\tau)=&\int \mathrm{d}\mathbf{p}^{\prime}%
\mathrm{d}\mathbf{p}^{\prime\prime}\delta\left(  \tfrac{{\mathrm{p}^{\prime}%
}^{2}-{\mathrm{p}^{\prime\prime}}^{2}}{2m}\right)  \frac{f_{\mathbf{p}_{0}%
}(\mathbf{r}^{\prime},\mathbf{r}^{\prime\prime},\mathbf{r},\mathbf{p}^{\prime
})}{2\pi m\mathcal{A}}\nonumber\\
=&\frac{1}{\mathcal{A}}\int\mathrm{d}\mathbf{p}^{\prime}\ f_{\mathbf{p}_{0}%
}(\mathbf{r}^{\prime},\mathbf{r}^{\prime\prime},\mathbf{r},\mathbf{p}^{\prime
}).
\end{align}
Applying this general procedure to the function $f_{\mathbf{p}_{0}}$ of
Eq.~(\ref{eq:TGF2}) we have%
\begin{align}
\mathcal{F}_{\mathbf{p}_{0}}(\mathbf{r},t)=&\frac{\sigma\Delta T}%
{2\pi^{3/2}\hbar^{2}\mathcal{A}}\int\mathrm{d}\mathbf{p}^{\prime}\exp\left[
\tfrac{\mathrm{i}}{\hbar}\mathbf{p}^{\prime}\cdot(\mathbf{r}-\mathbf{r}%
_{0})\right. \nonumber\\
& \left.  -\tfrac{\mathrm{i}}{\hbar}\frac{{\mathrm{p}^{\prime}}^{2}}%
{2m}t-\tfrac{\sigma^{2}}{2\hbar^{2}}(\mathbf{p}^{\prime}-\mathbf{p}_{0}%
)^{2}\right]  ,
\end{align}
since the integral over $\tau$ is now trivial. Performing the Gaussian
integral over $\mathbf{p}^{\prime}$ we obtain a wave-packet that refocalizes
with the same shape of the original one, but with momentum $-\mathbf{p}_{0}$.
The magnitude of the signal close to the maximum refocalization condition is
given by%
\begin{equation}
\left\vert \mathcal{F}_{\mathbf{p}_{0}}(\mathbf{r},t)\right\vert =
\frac{\mathcal{F}_{\mathrm{max}}}{\sqrt{1+(\mathcal{V}t/\sqrt{2}\sigma)^{2}}}
\exp{\left[  -\frac{\left(  \mathbf{r}-\mathbf{r}_{0}+\frac{\mathbf{p}_{0}}%
{m}t\right)  ^{2}}{2\sigma^{2}+(\mathcal{V}t)^{2}}\right] }.\label{eq:F_p0}%
\end{equation}

Numerical calculations of time-reversal imaging have been performed in
Ref.~\cite{cit-Draeger} for a two-dimensional cavity with the shape of a
sliced disk. The signal reconstruction could be visualized and a qualitative
agreement with the experimental results was found. Since we dispose now of a
quantitative semiclassical theory of refocusing it is important to test our
predictions in a stadium billiard, which is a paradigm of classical chaotic
dynamics. We calculate the evolution of the wave-packet through a second order
Trotter-Suzuki algorithm for a discrete Schr\"{o}dinger equation. Lattice
effects are minimized by considering $a_{0}\ll\lambda_{\mathrm{B}}\ll\sigma\ll
L_{b}$, where $a_{0}$ is the lattice constant, $\lambda_{\mathrm{B}}$ the de
Broglie wave-length associated with $\mathbf{p}_{0}$, and $L_{b}$ the size of
the billiard. We assume that at injection time all the original signal has
already decayed.

In Fig.~\ref{fig-1}(b) we show the numerical results for the time dependence
on the reconstructed signal at $\mathbf{r}_{0}$. The normalized $\mathcal{F}%
\mathcal{A}/\Delta T$ are well described by the semiclassical prediction
(thick solid) confirming the scaling with $\Delta T$ and $\mathcal{A}$ of
Eq.~(\ref{eq:F_m}). The normalizing factor for the numerical results is
approximately 1.4 times the semiclassical one. Such a difference may be due to
our discretization of the quantum problem as well as the difficulties of the
diagonal approximation to recover exact numerical values. The signal-to-noise
ratio does not change appreciably when the recording time is doubled, while it
is improved by increasing $\mathcal{A}$. In the right inset we show the
spatial reconstruction of the wave-packet around $\mathbf{r}_{0}$ for the
refocusing time $t=0$. We see that the semiclassical prediction (thick solid)
provides the proper scaling behavior and, up to the normalization factor, a
quantitative description of the TRM results.

The numerical implementation of TRM for integrable geometries results in a
refocusing that strongly depends on the position of the transducers and with a
signal hardly distinguishable from the background. The semiclassical approach
allows to understand this important difference between chaotic and integrable
systems. In the former the exponential proliferation of trajectories allows to
encode the information at all times, while in the latter the registered signal
will be strongly dependent on whether or not the source and the transducer are
connected by a stable trajectory.

Experimentally, the TRM procedure has been shown to be robust against local
and global perturbations introduced between the recording and injection phases
\cite{cit-Derode}. Even in the absence of these perturbations, it is natural
to expect that in any TRM setup the environment acting during the re-emission
is slightly modified respect to that of the recording phase. In the same
spirit of the LE studies, we can model this situation by assuming that in the
recording process we have a Hamiltonian $H$ that determines $G(\mathbf{r}%
_{i},\mathbf{r}^{\prime},\tau)$ in Eq.~(\ref{eq_twoGF}), while a modified
$\tilde{G}(\mathbf{r},\mathbf{r}_{i},t+\tau)$ is governed by the slightly
different Hamiltonian $\tilde{H}$. For the LE the details of the perturbation
$\tilde{H}-H$ are not important, and its effect can be accounted, after some
averaging, by affecting the contribution of each trajectory $s$ by an
additional factor $\exp[-L_{\hat{s}}/2\tilde{l}]$, where $\tilde{l}$ is an
effective mean-free-path associated to the perturbation. In general $\tilde
{l}$ depends on the velocity of the particle, i.e. for perturbations modeled
by an auxiliary impurity potential characterized by a strength $\gamma$ we
have $1/\tilde{l}=\gamma\tau^{2}/L_{\hat{s}}^{2}$ \cite{cit-Jalabert}.
Including this $\tau$-dependent phase prevents us of using the ergodic
approach, but working along the lines of Eqs.~(\ref{eq:F_m-inter}%
)-(\ref{eq:F_time}) we obtain a maximum refocusing for a non-static
environment given by%
\begin{align}
\mathcal{F}_{\mathrm{max}}(\gamma)=&\frac{2\mathcal{V}}{\sqrt{\pi}%
\sigma\mathcal{A}\gamma}\int_{0}^{\infty}\mathrm{d}\eta\ \eta^{2}\exp
[-\eta^{2}]\nonumber\\
& \times\left(  \exp\left[  -\frac{\gamma L_{1}}{\eta\mathcal{V}^{2}}\right]
-\exp\left[  -\frac{\gamma L_{2}}{\eta\mathcal{V}^{2}}\right]  \right)
,\label{eq:F_m_gamma}%
\end{align}
which reduces to $\mathcal{F}_{\mathrm{max}}\left(  1-(t_{2}+t_{1}%
)/4\tilde{\tau}\right)  $ for $t_{2}~\ll~\tilde{\tau}$; to $\mathcal{F}%
_{\mathrm{max}}\exp{\left[  -(t_{2}+t_{1})/4\tilde{\tau}\right]  }$ for
$t_{2}\!-\!t_{1}~\ll~\tilde{\tau}$; and to $c\tilde{\tau}/(\sqrt{\pi}%
\sigma\mathcal{A})\exp{\left[  -c^{\prime}t_{1}/2\tilde{\tau}\right]  }$ for
$t_{2}\!>\!\tilde{\tau}$. The numerical constants $c=2.94$ and $c^{\prime
}=0.46$ are given by rational-argument values of the $\Gamma$ function, and
the characteristic time is defined as $\tilde{\tau}=\mathcal{V}/(2\sqrt{\pi
}\gamma)$. The proportionality of the reconstructed signal with the injection
interval is clearly lost in the limit of $t_{2}\!>\!\tilde{\tau}$ since the
perturbation renders ineffective the longest recording times. From the
relevant limits that we have singled out, the second one is the most important
for current experiments. It has a Fermi-Golden-Rule structure that can be
obtained under very general considerations. For perturbations where the
effective elastic mean-free-path $\tilde{l}$ increases with the Lyapunov
exponent of the unperturbed system (i.e. mass distortion in a Lorentz gas
\cite{cit-Jalabert}) the characteristic time $\tilde{\tau}$ increases with the
chaoticity of the system. Similarly to the Fermi-Golden-Rule regime of the LE,
such a motional narrowing effect translates into larger stability, and
improved refocalization, for the more chaotic systems, in agreement with the
experimental findings \cite{cit-Derode}.

In summary, we have described the refocalization signal for the time reversal
mirror procedure through the semiclassical approximation. The chaotic nature
of the underlying classical dynamics appears as a key ingredient to ensure the
stability of the refocalization towards perturbations and the proportionality
of the recovered signal with the injection time.

We thank T. Kramer and K. Richter for valuable discussions and A. Goussev for
helpful correspondence. We acknowledge financial support of the
French-Argentinian program ECOS-Sud and the Deutsche Forschungsgemeinschaft
within FG 760. RAJ acknowledges the hospitality of the Universit\"at
Regensburg and HMP that of the ICTP of Trieste and the MPI-PKS of Dresden.

\end{document}